\documentclass[aps,pre,reprint]{revtex4-2}
\usepackage{amsmath,amssymb,graphicx,bm,hyperref,float,subcaption,xcolor,float,nicefrac}
\usepackage[justification=raggedright,singlelinecheck=false]{caption}
\date{\today}

\begin{document}
	\title{Conductance fluctuations in random resistor networks with hyperuniform disorder}
	\author{Bikram Pal}
	\email{bikram25052005@gmail.com}
	\address{Department of Physical Sciences, Indian Institute of Science Education and Research Kolkata, Mohanpur, 741246, India}
	\address{International Centre for Theoretical Sciences, Tata Institute of Fundamental Research, Bengaluru 560089, India}
	\begin{abstract}
		We study conductance fluctuations in random resistor networks with hyperuniform bond disorder, where the fluctuations of the number of bonds present in a test volume $V$ scale as $V^{-a}$ with $a > 1/2$. Since small changes in the concentration of bonds present in a local region give rise to a proportionate increase in the locally averaged conductance, one may expect that in hyperuniform disorder, conductance fluctuations will also show suppressed fluctuations. We argue that this is not the case: conductance fluctuations scale as $L^{-d/2}$ for a sampling size $L$. We show numerical results for $d=2$.
	\end{abstract}
	\maketitle
	\section{Introduction}
	\label{intro}
	
	The conductivity of disordered random media is a problem of fundamental interest. A standard class of simple, tractable models for disordered systems is given by electrical networks consisting of resistors with randomness in the values of resistances \cite{percolation1, percolation2, percolation3}. Clearly, there is no conduction if the concentration of bonds present is below the percolation threshold ($p_c$). Historically, considerable effort has been devoted to understand the universal behavior of conductance fluctuations near the percolation threshold \cite{Kirkpatrick1971, Kirkpatrick1973, HarrisFisch1977, Bergman1981, Bernasconi, Shklovsk}.
	
	Numerical simulations have shown that the conductance follows a power-law dependence characterized by $\sigma \sim (p-p_c)^t, \,\, (p>p_c)$ where $\sigma$ is conductance and $t$ is conductance exponent that depend on the spatial dimension \cite{Straley1977,WebmanJortnerCohen1977}. It was noted by de Gennes that the equations governing the conductance of a disordered resistor network are equivalent to those encountered in determining the bulk elastic coefficients of a random spring network \cite{deGennes1976}. Later, in the 1980s, Derrida et al. studied this problem near the percolation threshold using a transfer matrix formalism \cite{DerridaVannimenus1982}. Batrouni et al. studied the broad distribution of sizes of jumps in conductance near critical point \cite{BatrouniRedner1988}. The critical behavior of resistance fluctuations using the links-nodes-blobs picture of the percolating cluster were analyzed and shown that the relative resistance fluctuations diverge as the system approaches $p_c$ \cite{WrightBergmanKantor1986}. In this paper, our interest is the behavior of off-critical conductance fluctuations.
	
	Understanding the current distribution in random resistor networks has long been an actively investigated problem. Near $p_c$, it was found that these distributions are broad, with different moments scaling with different powers of the system size, and hence called multifractal. \cite{Aharony1992, deArcangelisRednerConiglio1986, Kolek2000}. Green's function based methods have enabled systematic calculations of transport in networks with a finite number of modified bonds, yielding exact results in both the weak and strong-disorder regimes \cite{BhattacharjeeRamola2023}.
	
	In equilibrium systems with uniform density and finite compressibility, it is well known that density fluctuations within a small volume $V$ scale as $V^{-1/2}$. Recently, significant attention has been directed toward systems in which density fluctuations are suppressed and instead scale as $ V^{-a} $, with $ a > 1/2 $ and correspond to the system having a zero value of thermodynamic compressibility. Such systems are called \textit{hyperuniform} \cite{TorquatoRev1,TorquatoRev2}. These systems have attracted considerable interest in recent years as they arise in a broad range of condensed-matter and solid-state systems, with applications including amorphous semiconductors, photonic materials, and other disordered media \cite{ice,silica,bird,Indranil1,Indranil2}. In this work, we investigate the conductance and its fluctuations in systems exhibiting hyperuniform bond disorder. The problem is nontrivial because current fluctuations in resistor networks display long-ranged correlations, a consequence of current conservation within the network except at the boundaries.
	%%%%%%%%%%%%%%%%%%%%%%% Current and Voltage Distribution %%%%%%%%%%%%%%%%%%%%%%%%%%%%%%%%%%%%%%%%%%%%
	\begin{figure*}[t]
		%\centering
		\includegraphics[width=1.05\linewidth]{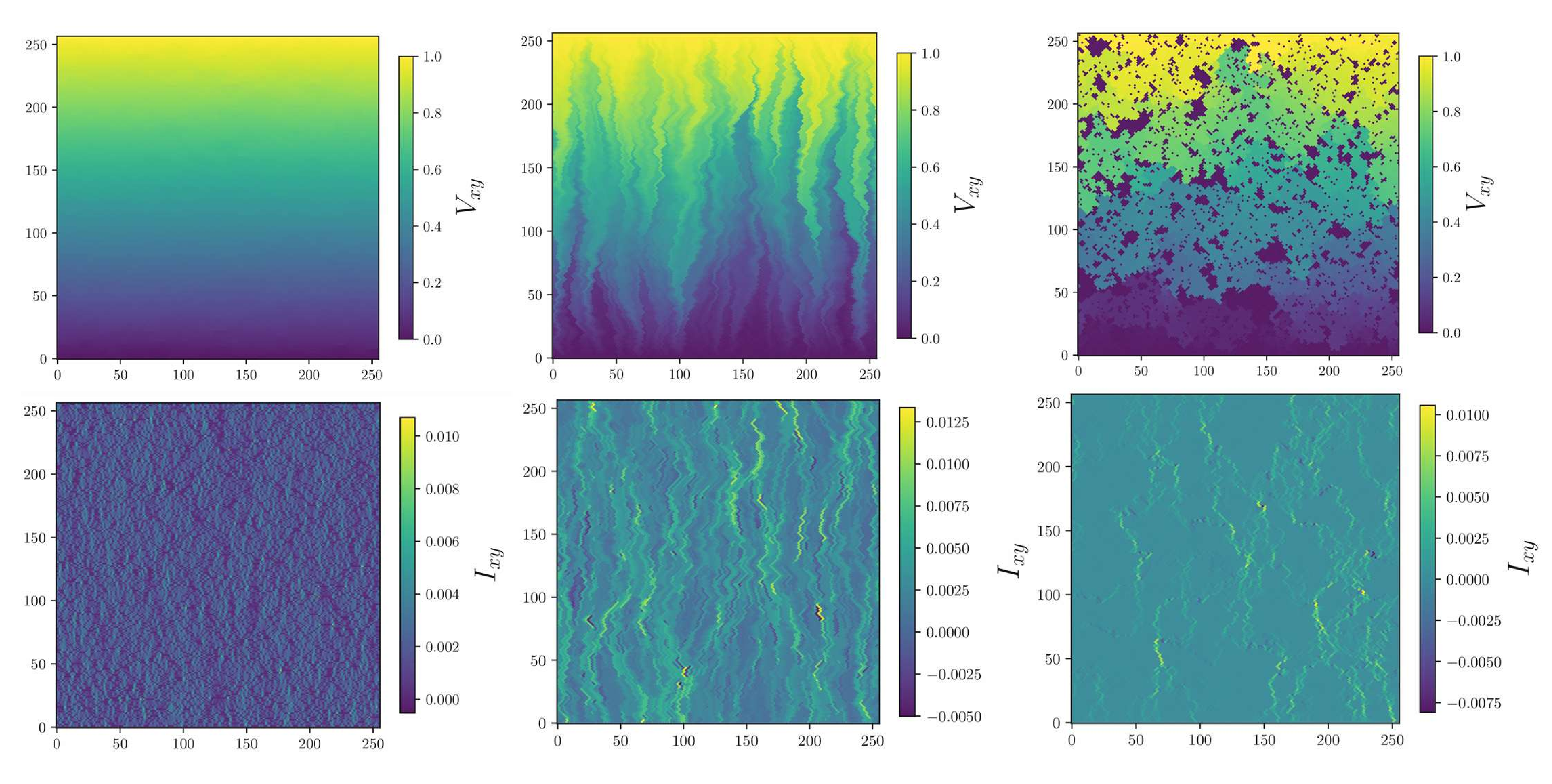}
		\caption{The plots above show the spatial distribution of potential (top row) and current (bottom row) for one realization of different models with $L=256$. From left to right (in both rows) we have Model A ($q=1$), Model B and Model C respectively.}
		\label{fig:space}
	\end{figure*}
	%%%%%%%%%%%%%%%%%%%%%%%%%%%%%%%%%%%%%%%%%%%%%%%%%%%%%%%%%%%%%%%%%%%%%%%%%%%%%%%%%%%%%%%%%%%%%%%%%%%%%
	
	Consider a $d$-dimensional system with bond percolation disorder, in which a fraction $p$ of the bonds are present. Each occupied bond has unit resistance, while absent bonds are assigned infinite resistance. Let $\sigma^*(p)$ denote the bulk conductance in the thermodynamic limit. A natural approach to analyze such a system is to coarse-grain it over a length scale $\ell$, with $\ell \ll L$ where $L$ is the system size, and define an effective conductance $\sigma$ as a function of the coarse-grained bond concentration $c$. For a small variation in the concentration one might expect
	\[\delta \sigma \approx \frac{d\sigma^*}{dc}\Big|_{c}\,\delta c .\]
	This argument would suggest that if concentration fluctuations at scale $\ell$ vary as $\ell^{-a}$, with $a>d/2$, then conductance fluctuations should also exhibit hyperuniform scaling, varying as $\ell^{-a}$. However, this naive expectation is not correct. The reason is that current fluctuations in the network exhibit long-ranged correlations, whereas concentration fluctuations in different regions are independent in the standard definition of percolation disorder. For example, Fig.~\ref{fig:space} shows the current distribution in different models of random resistor networks (defined later in section~\ref{model}), where the long-ranged correlations in the current fluctuations are clearly evident.
	
	Given the strong correlations of current fluctuations in the longitudinal direction (along the bulk flow direction), one may ask whether the current correlations across a transverse surface of section are only short-ranged. If that were the case, the variance of the net current across a section of linear size $L$ would scale as $L^{-(d-1)}$. This would imply conductance fluctuations scaling as $L^{-(d-1)/2}$. We will see below that even this expectation is not correct.
	
	Lastly, it is well known that in resistance measurements in disordered materials, a major contribution to the measured resistance may arise from the contact resistance at the boundaries. If the bulk conductance fluctuations were dominated by disorder in the boundary region, one would again expect the variance of the bulk conductance to scale as $L^{-(d-1)/2}$. We will argue below that this scenario is also not realized. In fact, we will present numerical evidence (for $d=2$) that variance of conductance varies as $L^{-d}$.
	
	This paper is organized as follows. In section~\ref{model} we describe our definition of models and construction. In section~\ref{result} we present our results, plots and a perturbative explanation near $p=1$. In section~\ref{discussion} we present our discussion.
	
	\section{Model Definition}
	\label{model}
	We construct a class of bond diluted random square lattice where each site has co-ordination number $2$ or more. A bond, if present, is assigned unit resistance, whereas an absent bond corresponds to infinite resistance. One of the models of this class can be constructed using full dimer coverings. We remove $q$ fraction $(0\leq q\leq 1)$ of the dimer-ed bonds from the lattice to make a lattice with co-ordination number $3$ or more. At $q=1$ every bulk site has exactly three neighboring bonds. Since the number of bonds $N_b$ in a region containing $N_s$ sites satisfies $2N_b = 3N_s$. Thus the fluctuations of $N_b$ within a window arise only from boundary contributions rather than bulk randomness. Consequently $\mathrm{Var}[N_b(L)] \sim L^{d-1}$ instead of $L^d$, and the variance of the bond density vanishes as $L \to \infty$. Hence the bond distribution is hyperuniform when $q=1$.
	
\begin{figure*}[t]
		\centering
		\includegraphics[width=0.75\textwidth]{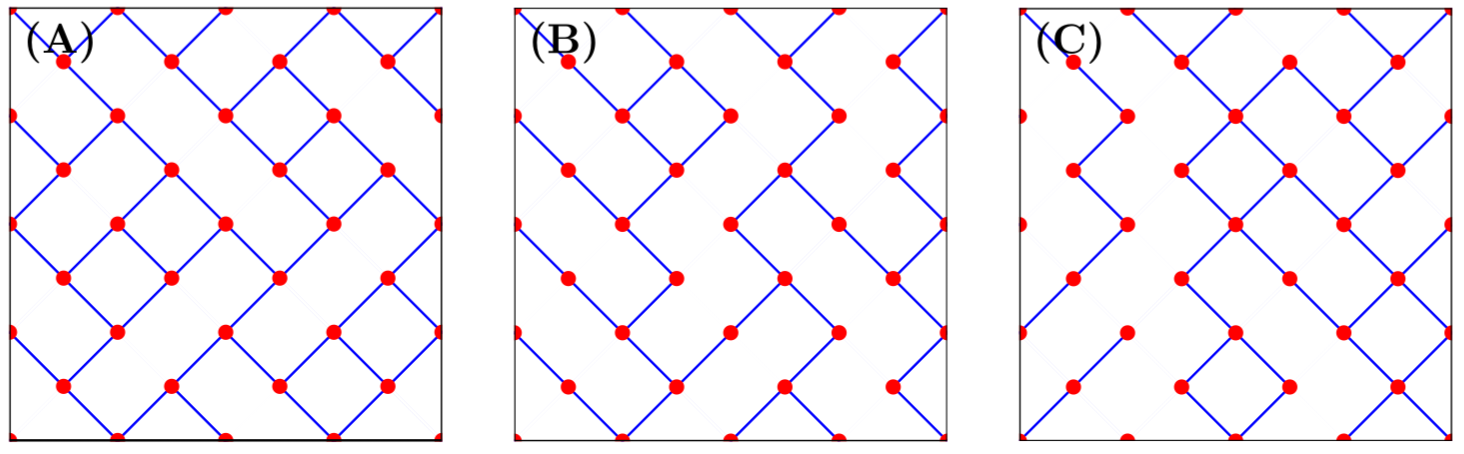}
		\caption{From left to right, (A) shows model A with $q=1$, (B) shows model B and (C) shows model C.}
		\label{fig:lattice}
	\end{figure*}

	To make such a lattice, we uniformly sample full dimer coverings on lattice using an ergodic worm algorithm that flips dimers along self-avoiding closed loops, ensuring detailed balance and uniform equilibrium distribution, following ref.\cite{Rakala2021}. We call this \textit{Model A}. Clearly, such a lattice belongs to the infinite cluster. An example of this is shown in Fig.~\ref{fig:lattice}.
	
	%%%%%%%%%%%%%%%%%%%%%%% Model Figure %%%%%%%%%%%%%%%%%%%%%%%%%%%%%%%%%%%%%%%%%%%%%%%%%%%
	%%%%%%%%%%%%%%%%%%%%%%%%%%%%%%%%%%%%%%%%%%%%%%%%%%%%%%%%%%%%%%%%%%%%%%%%%%%%%%%%%%%%%%%%%
	
	For comparison, we define \textit{Model B} where we take a tilted lattice of size $L$ (as shown in Fig.~\ref{fig:lattice}) such that every site has at least one bond above and at least one bond below; and \textit{Model C} which is the standard percolating lattice away from $p_c$. In \textit{Model B} every site has co-ordination number $2$ or more. Moreover, Model B, like Model A, forms a single connected cluster, but does not exhibit hyperuniformity. In model C, the fractional number of sites in the infinite cluster is strictly less than $1$ for $p<1$. Given a value of $p$($>0.5$ for this case), we generate these lattices randomly by drawing cyclic binary chains of length $L$ with $p$ fraction of $1$s and no consecutive $0$s (and drop this condition to generate standard percolating lattice). 
	
	All models have periodic boundaries in $x$ and open boundaries in $y$ direction. We apply $L$ unit potential difference across the vertical direction of the lattice such that the current in each bond is $\mathcal{O}(1)$. Define bulk conductance denoted by $\sigma$ such that it is normalized (i.e., $\sigma=1$ when all bonds are present). To determine $\sigma$ we solve Kirchhoff's equation \cite{Kirchhoff1847} at each site $(x,y)$:
	$$\sum_{(x,y)\sim(x',y')}g_{(x,y),(x',y')}\left(V(x,y)-V(x',y')\right)=0$$
	where $V(x,y)$ denotes potential at site $(x,y)$ and $g$ denotes conductance of the bonds which takes values $0$ or $1$ depending on whether the bond is present or not. Given $(L,p)$, for many independent realizations, we obtain a probability distribution $F(\sigma)$ (with a probability density $\mathcal{P}(\sigma)$) and we want to find how the width of $\mathcal{P}(\sigma)$ scales as $L$.

	\section{Results}
	\label{result}
	%%%%%%%%%%%%%%%%%%%%%%%%%%%%%%%%%% Main Figures %%%%%%%%%%%%%%%%%%%%%%%%%%%%%%%%%%%%%%%%%%%%%%%%%%%%%%
	\begin{figure*}
		
		% First row 
		\begin{subfigure}{0.5\textwidth}
			\centering
			\includegraphics[width=0.95\linewidth]{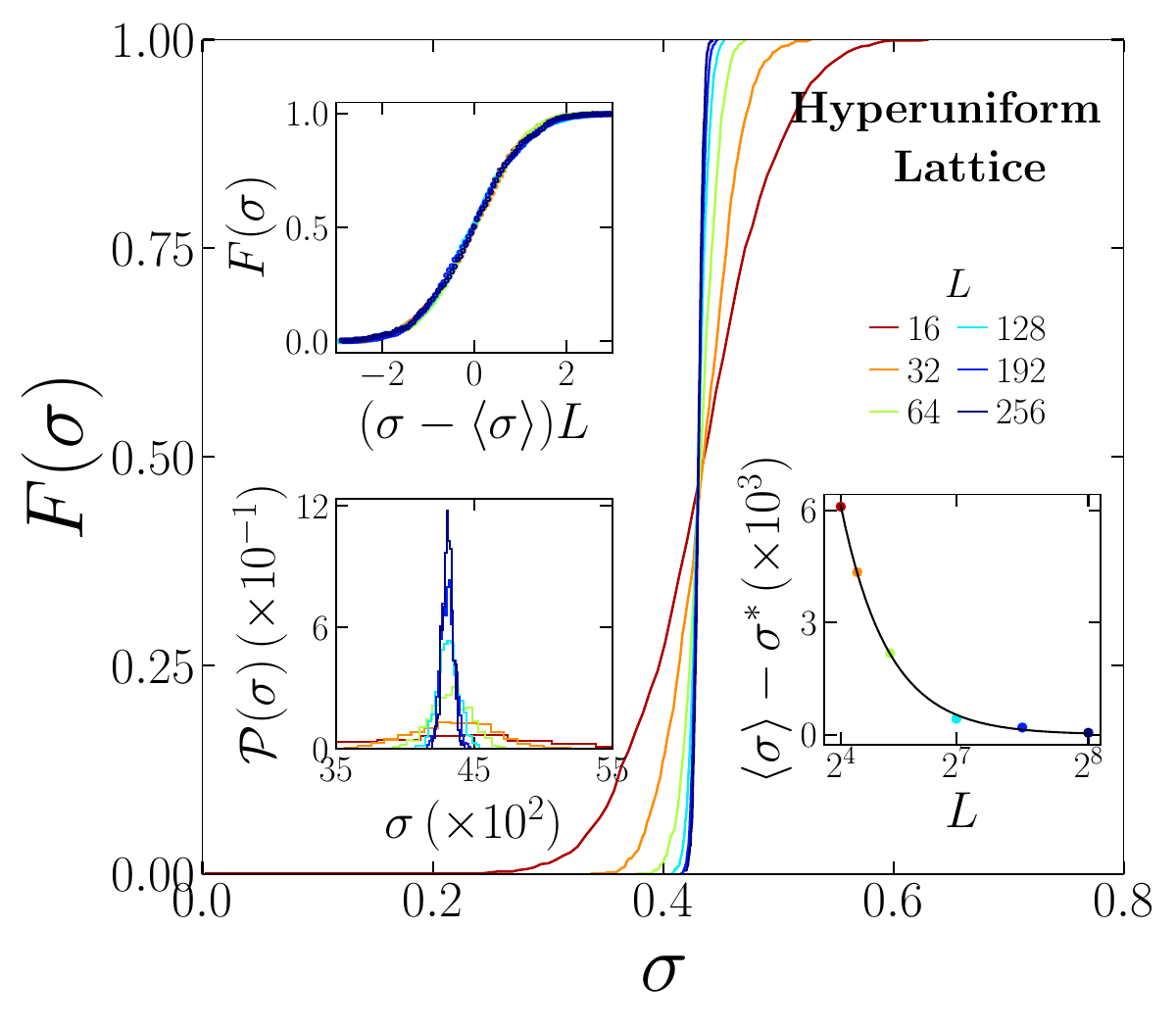}
		\end{subfigure}\hfill
		\begin{subfigure}{0.5\textwidth}
			\centering
			\includegraphics[width=0.95\linewidth]{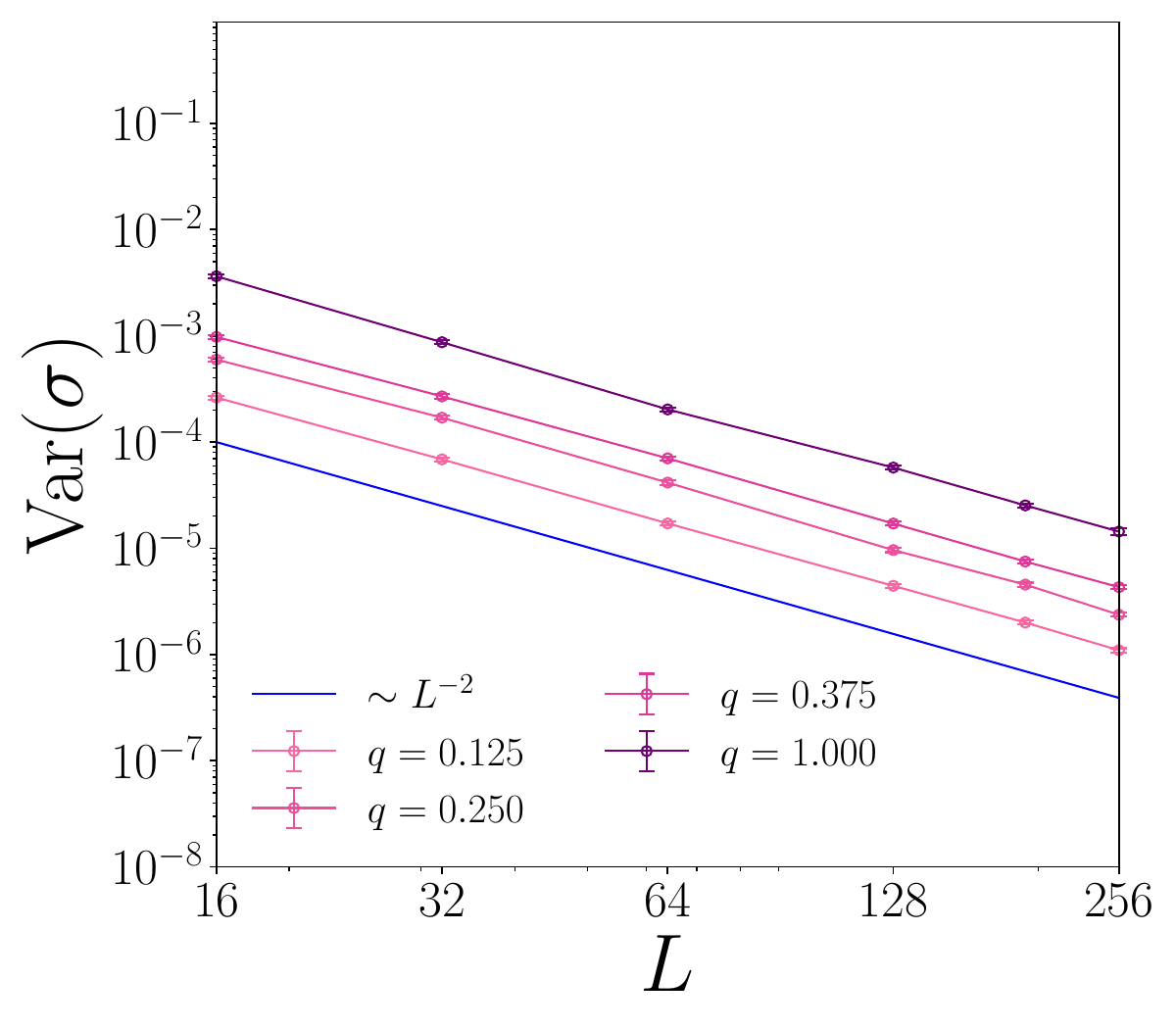}
		\end{subfigure}
		% Second row 
		\begin{subfigure}{0.5\textwidth}
			\centering
			\includegraphics[width=0.95\linewidth]{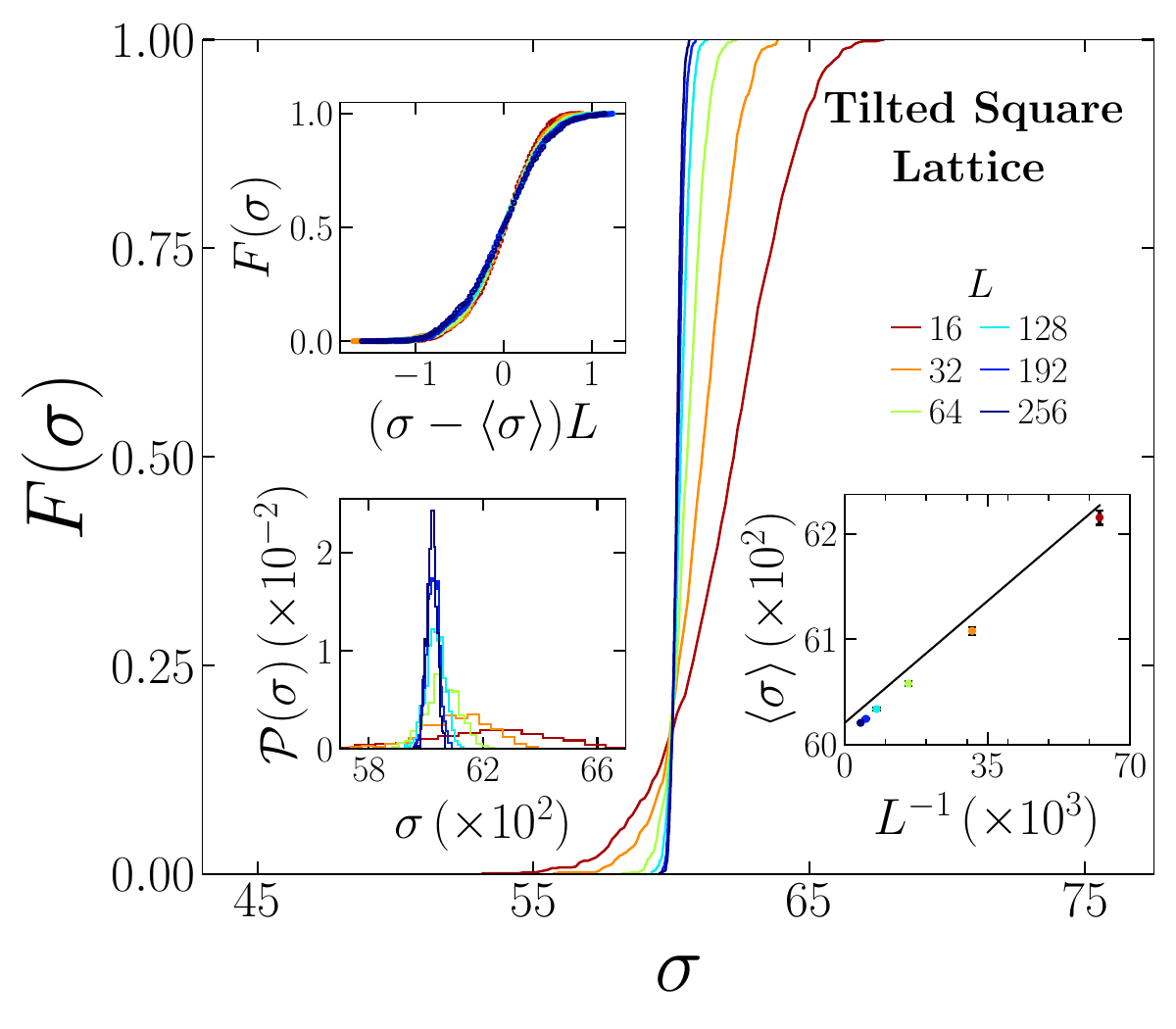}
		\end{subfigure}\hfill
		\begin{subfigure}{0.5\textwidth}
			\centering
			\includegraphics[width=0.95\linewidth]{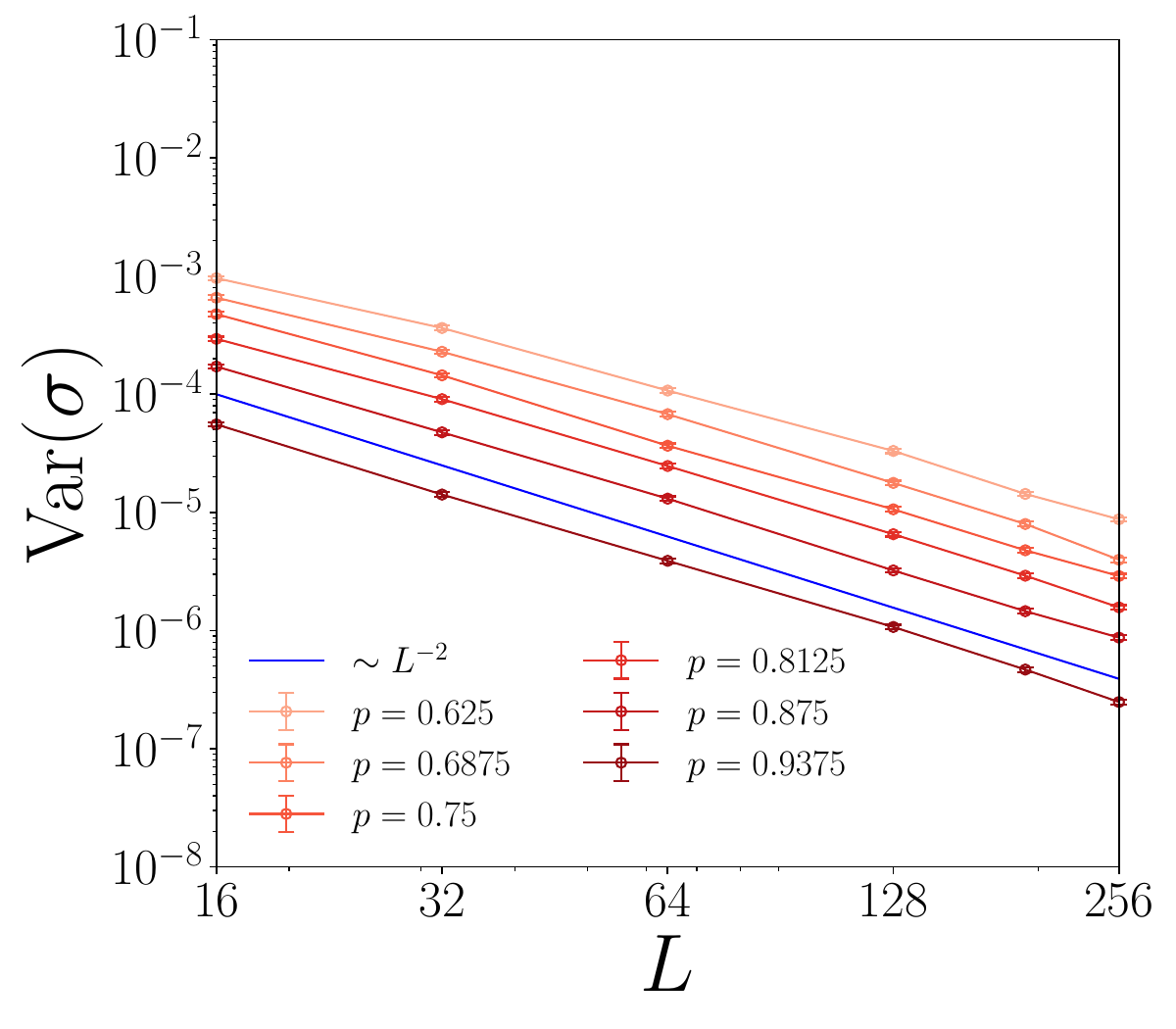}
		\end{subfigure}
		% Third row 
		\begin{subfigure}{0.5\textwidth}
			\centering
			\includegraphics[width=\linewidth]{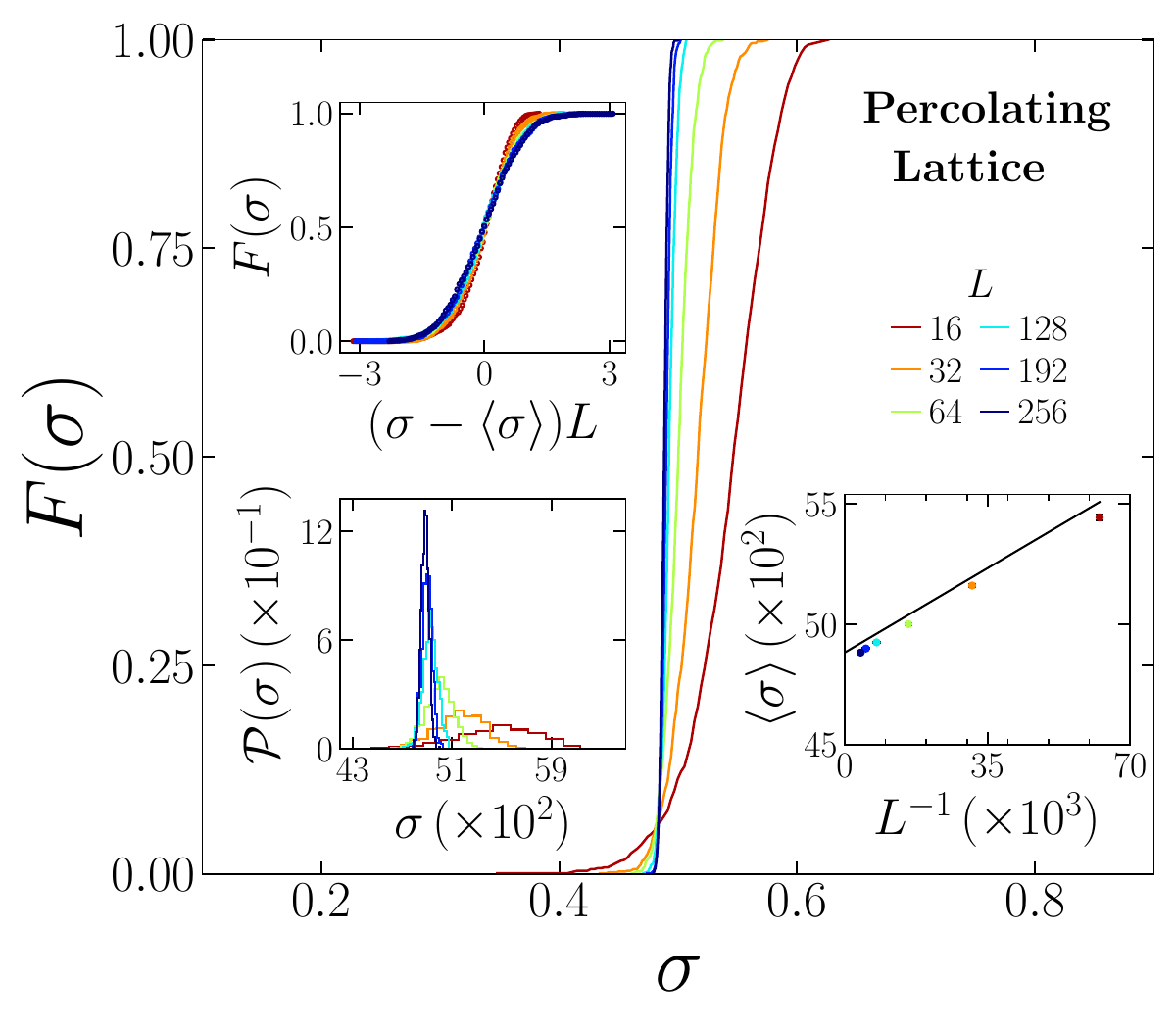}
		\end{subfigure}\hfill
		\begin{subfigure}{0.5\textwidth}
			\centering
			\includegraphics[width=0.97\linewidth]{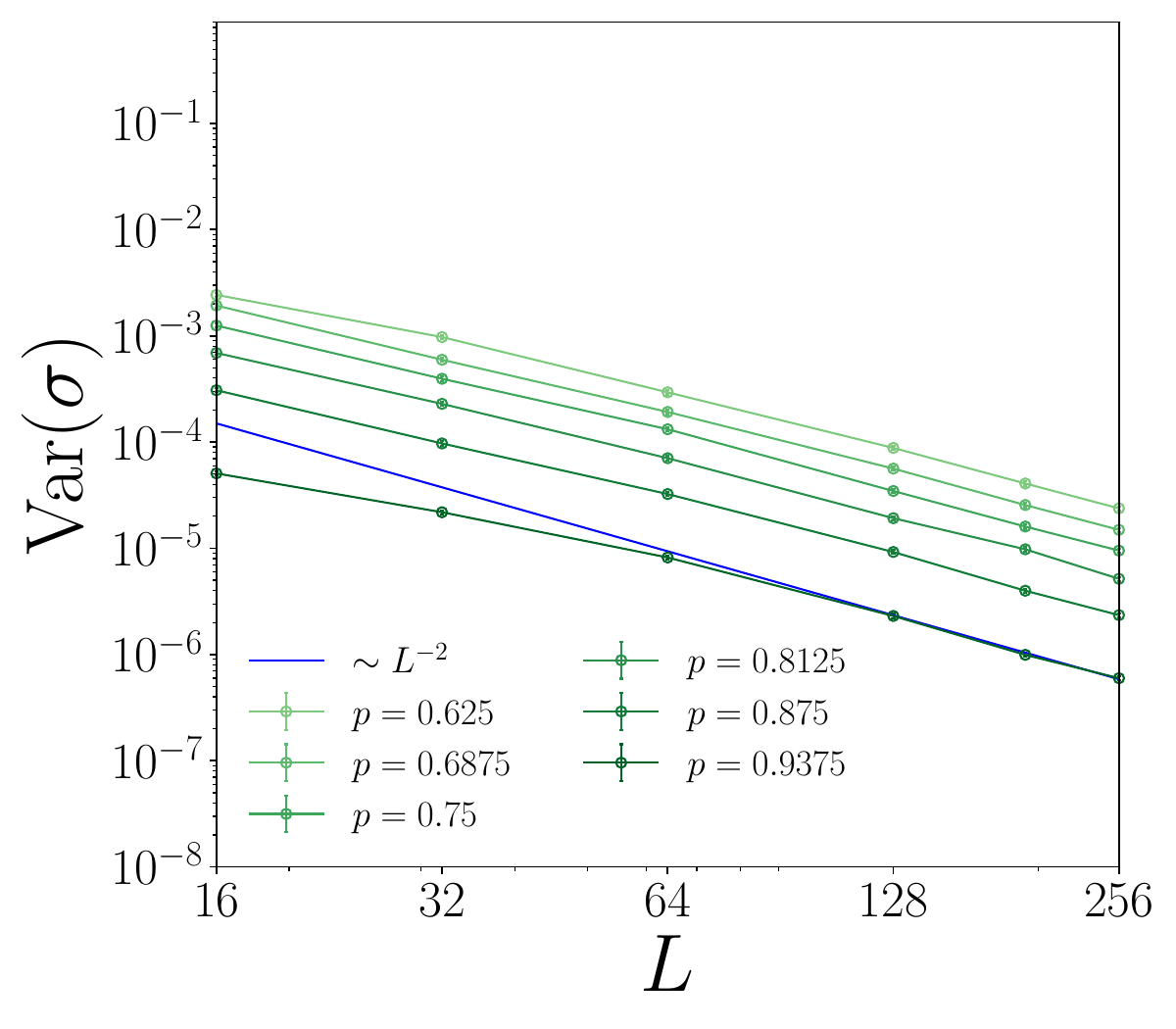}
		\end{subfigure}
		\caption{From top to bottom we have Model A with $q=1$, Model B with $p=0.75$ and Model C with $p=0.75$ respectively. $F(\sigma)$ denotes cumulative distribution of $\sigma$ and $\mathcal{P}(\sigma)$ denotes probability density.}
		\label{fig:pdf}
	\end{figure*}
	%%%%%%%%%%%%%%%%%%%%%%%%%%%%%%%%%%%%%%%%%%%%%%%%%%%%%%%%%%%%%%%%%%%%%%%%%%%%%%%%%%%%%%%%%%%%%%%%%%%%
	
	Fig.~\ref{fig:pdf} shows the probability distribution $F(\sigma)$ and probability density $\mathcal{P}(\sigma)$ of conductance. We show that the $F(\sigma)$ for different $L$ shows a scaling scaling collapse if plotted as a function of $L\left(\sigma-\langle\sigma\rangle\right)$ (refer to insets of Fig.~\ref{fig:pdf}). All of them have effectively $75\%$(for hyperuniform case $q=1$) bonds present in the network. Clearly, as $L$ increases the density function shifts left and the curve sharpens. This shift is more pronounced in Model B and C as compared to $q=1$ of Model A which has hyperuniformity. Away from critical point, we find that for model B and C, (refer to insets of Fig.~\ref{fig:pdf})
	\[\langle\sigma(L,p)\rangle=\sigma^*+\frac{a}{L}+\,\mathrm{higher}\,\,\mathrm{order}\,\,\mathrm{terms}\,\,\mathrm{in}\,\,\frac{1}{L},\]
	where $\langle.\rangle$ denotes ensemble average, $a>0$. On the other hand, the shift of mean for hyperuniform case is found to be exponentially decaying
	\[\langle\sigma(L,p)\rangle=\sigma^*+be^{-cL}+\cdots,\]
	where $b,c>0$. For the hyperuniform network we find that, (where effectively $75\%$ bonds are present) $\sigma^*= 0.43075(6)$ which is smaller than that of the standard percolation lattice (Model C with $p=0.75$) $\sigma^*=0.4846(1)$ and even lesser than Model B (with $p=0.75$) $\sigma^*=0.60082(3)$. However, our main interest is in understanding the sample to sample fluctuations, and not in the ensemble averaged mean value. 
	We find the variance of conductance scales as
	\[\mathrm{Var}(\sigma)\sim L^{-2}.\]
	At present we do not have a rigorous derivation of this result. Instead, we present numerical evidence and a heuristic argument that help clarify its origin.
	
	%%%%%%%%%%%%%%%%%%%%%%%% Cdelta scatter %%%%%%%%%%%%%%%%%%%%%%%%%%%%%%%%%%%%%%%%%%%%
	\begin{figure*}[t]
		\includegraphics[width=0.425\linewidth]{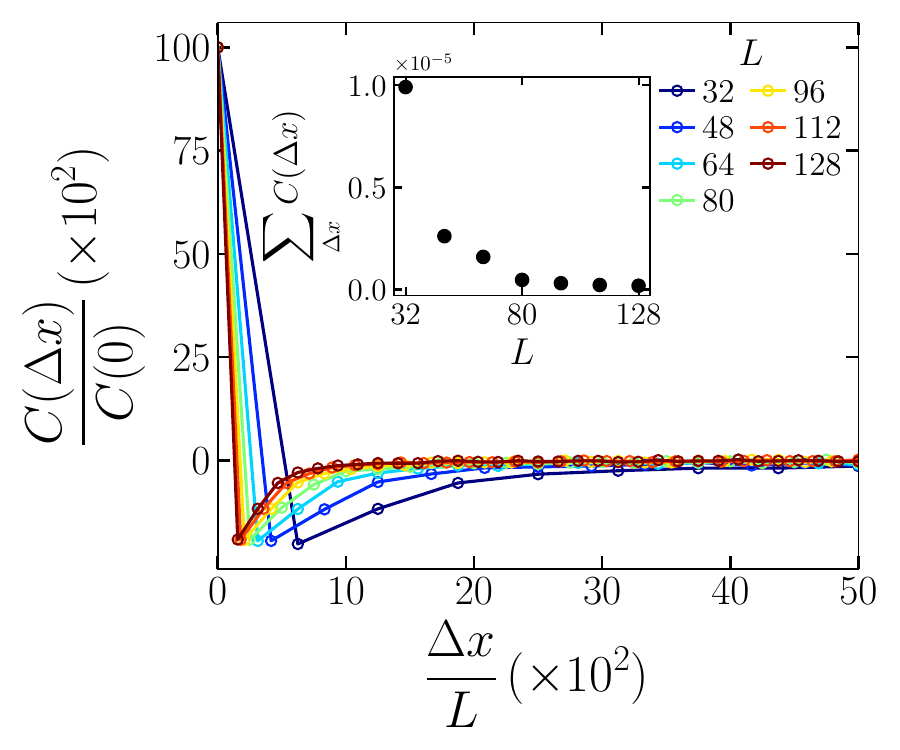}
		\includegraphics[width=0.365\linewidth]{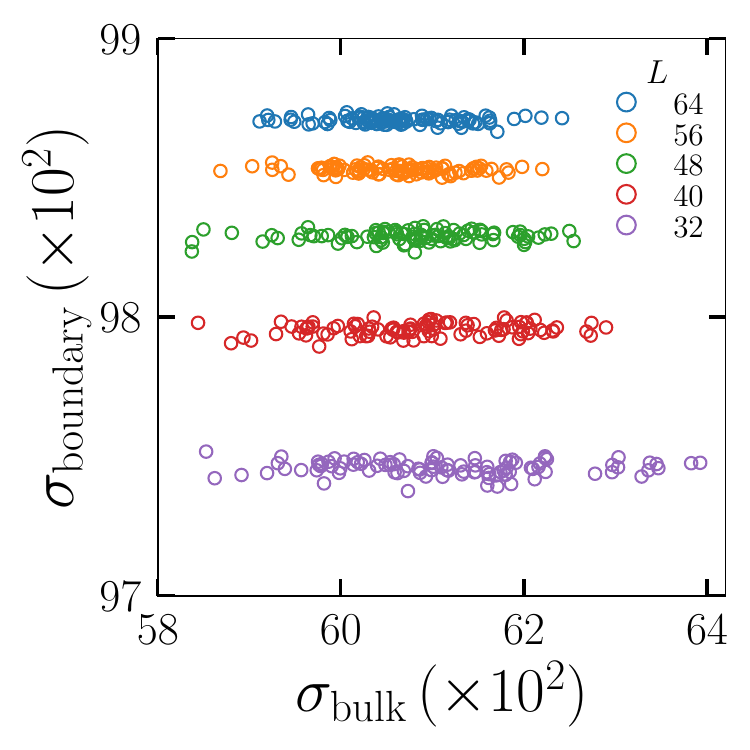}
		\caption{(Left) Plot of current-current correlation function $C(\Delta x)$ versus $\Delta x/L$.
			(Right) Scatter plot of conductance when disorder was in the bulk versus when only restricted to the boundaries.}
		\label{fig:Candscatter}
	\end{figure*}
	%%%%%%%%%%%%%%%%%%%%%%%%%%%%%%%%%%%%%%%%%%%%%%%%%%%%%%%%%%%%%%%%%%%%%%%%%%%%%%%%%%%%%%%%%%%%%%%%%%%%%
	
	The conductance can be written as the sum of local currents crossing any horizontal cross section of the system upto a normalization such that it is $\mathcal{O}(1)$,
	$$\sigma =\frac{1}{L} \sum_x I(x,y),$$
	where $I(x,y)$ denotes the current entering (or leaving) site $(x,y)$. Because of Kirchhoff’s current conservation law, the total current crossing any horizontal slice is independent of $y$ in the steady state. Thus fluctuations of the global conductance can be analyzed in terms of fluctuations of the local currents at a constant value of the $y$ coordinate. Current fluctuations exhibit strong correlations along the direction of current flow (the $y$ direction), but correlations along the transverse direction decay rapidly. Generically, this would lead to fluctuations that scales as $L^{-\nicefrac{1}{2}}$ for a sum over $L$ sites unless the integral of the correlation function vanishes. The variance of conductance can be written as
	\[Var(\sigma) =\frac{1}{L^2}\left[\sum_x Var(I(x,y)) +\sum_{x\neq x'} Cov(I(x,y),I(x',y))\right].\]
	Let $\delta I(x,y)$ denote the fluctuation of the current at position $x$. To understand the scaling of this expression, we examine the current--current correlation function
	\[C(\Delta x)=\langle \delta I(x,y)\,\delta I(x',y)\rangle ,\]
	where $\Delta x = x-x'$. Due to the translational invariance of the probability measure, the correlation function depends only on the separation, i.e. $C(\Delta x)=C(|x-x'|)$. The variance of the conductance can then be written as $\mathrm{Var}(\sigma)=A_L/L ,$ where
	\[A_L = \sum_{\Delta x} C(\Delta x).\]
	Clearly $A_L>0$ for all $L$. If the limit
	\[A=\lim_{L\to\infty} A_L\]
	is nonzero, then the variance would scale as $\mathrm{Var}(\sigma)\sim A/L$. Therefore, the variance can increase more slowly than $L^{-1}$ only if the correlation sum vanishes in the thermodynamic limit, i.e.
	\[\sum_{\Delta x} C(\Delta x)=0 .\]
	
	Fig.~\ref{fig:Candscatter} shows that $C(\Delta x)$ decays rapidly with separation along the $x$ direction, and our data indicate that the integral of the correlation function approaches zero as $L$ increases (see inset of Fig.~\ref{fig:Candscatter}). This implies that the leading $L^{-1}$ contribution cancels, leaving a subleading scaling $Var(\sigma)\sim L^{-2}$ which is supported by the fact that we see a good scaling collapse when $F(\sigma)$ is plotted against $(\sigma-\langle\sigma\rangle)L$.
	
	We now show evidence that surface fluctuations do not dominate bulk conductance. If this were not the case, the variance would scale as $L^{-(d-1)}$ in $d$ dimensions. To test this possibility we examined the correlation between conductance when we had bulk disorder and when disorder was only present in the boundary. As shown in Fig.~\ref{fig:Candscatter}, there is no significant correlation between these quantities. This indicates that surface fluctuations do not dominate the conductance fluctuations.
	
	Our numerical results show that such a cancellation indeed occurs, resulting in the observed scaling $Var(\sigma)\sim L^{-2}$. Thus the $L^{-2}$ scaling of conductance fluctuations is nontrivial.
	
	In $d$ dimensions the conductance can be written as a sum of local currents crossing a $(d-1)$-dimensional cross section, 
	$$\sigma=\frac{1}{L^{d-1}}\sum_{\mathbf{x}_\perp} i(\mathbf{x}_\perp).$$
	If current correlations along the transverse directions remain short ranged and the integral of the correlation function vanishes as observed in two dimensions, the leading contributions cancel in the same way. This would imply $Var(\sigma)\sim L^{-d}$, although a rigorous derivation remains an open problem.
	
\begin{figure}[t]
		\includegraphics[width=0.45\textwidth]{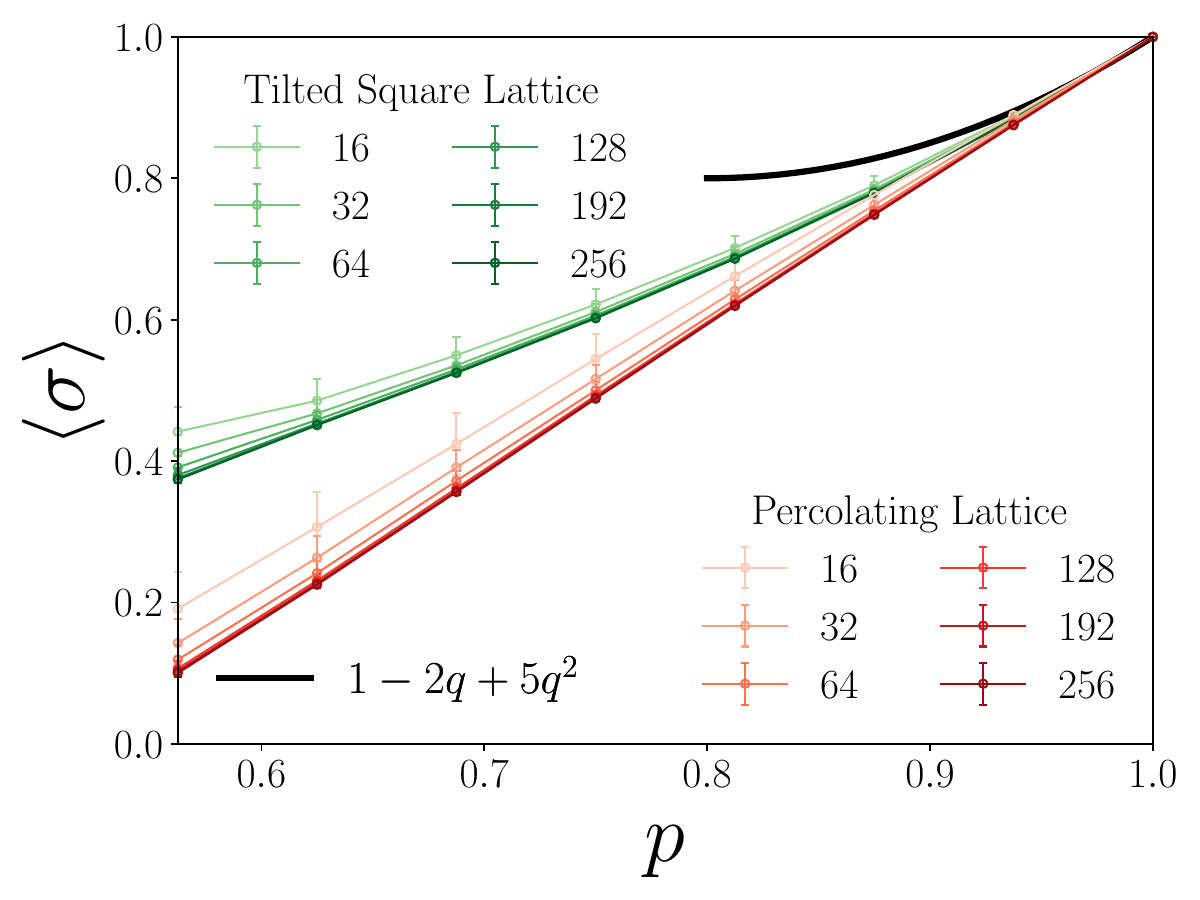}
		\caption{$\langle\sigma(p,L)\rangle$ is plotted as a function of $L$. The black line denotes the perturbative prediction near $p=1$.}
		\label{fig:sigma}
	\end{figure}

	Near the fully connected limit ($p=1$), the response of the resistor network to the removal of a small fraction $q=1-p$ of bonds can be treated using Green's function \cite{BhattacharjeeRamola2023, Cserti2002, Kang1994, Ravi1994}. For a single missing bond the electrostatic potential can be written as $\phi(x,y)=\phi^{(0)}(x,y)+\delta\phi(x,y),$ where $\phi^{(0)}=y/L$ is the uniform potential corresponding to homogeneous current flow, and $\delta\phi$ is the perturbation induced by the defect. The latter is equivalent to the potential generated by an effective dipole located at the removed bond and can be obtained from the lattice Green’s function. Because the governing equations are linear, the effect of multiple defects can be approximated as a superposition of single-bond perturbations. This motivates the expansion of the effective conductivity for small defect density $q$,
	\[
	\sigma(q)=1-c_1 q+c_2 q^{2}+\cdots .
	\]
	Comparison with numerical data gives $c_1=2$ and $c_2\simeq5$, yielding $\sigma(q)=1-2q+c_2 q^2$. This quadratic approximation remains accurate in the weak-disorder regime, deviating by less than $1\%$ even for $q=0.05$ (see Fig.~\ref{fig:sigma}).
	
	%%%%%%%%%%%%%%%%%%%%%%%%%%%%%%%%%%%%%%%%%%%%%%%%%%%%%%%%%%%%%%%%%%%%%%%%%%%%%%%%%%%%%%%%%%%%%%%%%%%%%%%%
	%%%%%%%%%%%%%%%%%%%%%%%%%%%%%%%%%%%%%%%%%%%%%%%%%%%%%%%%%%%%%%%%%%%%%%%%%%%%%%%%%%%%%%%%%%%%%%%%%%%%%%%%
	
	\section{Discussion}
	\label{discussion}
	
	In this work, we studied how hyperuniform disorder influences both the mean conductance and its fluctuations in random resistor networks. By comparing hyperuniform bond configurations with two reference models, we isolate the effect of suppressed long-wavelength density fluctuations on transport. Our main finding is that $\mathrm{Var}(\sigma)\sim L^{-2}$ in all three models, irrespective of whether the disorder is hyperuniform or not. In hyperuniform systems, the fluctuations in the fractional concentration of bonds in a system of size $L$ vary as $L^{-(d+1)/2}$. But current conservation induces long-ranged correlations in the current densities, and the fluctuations in the bulk conductance of a sample of size $L$ are  larger, and vary as $L^{-d/2}$, but not as large as $L^{-(d-1)/2}$, which would be valid if the current distribution within a transverse section had only short ranged fluctuations ($C(\Delta x)$ was short ranged), or if the bulk conductance fluctuations were dominated by the contact resistance fluctuations  at the ends. 
	
	While the asymptotic scaling of the variance is identical in all models, the approach to this scaling shows noticeable differences. In particular, the standard percolation lattice exhibits stronger finite-size effects, with the variance approaching its asymptotic $L^{-2}$ behavior more slowly than in the constrained tilted lattice and the hyperuniform network. This difference is consistent with the presence of larger structural fluctuations in the percolating lattice, which enhance finite-size corrections even though they do not modify the asymptotic scaling exponent.
	
	Our Green's function based numerical analysis suggests that the leading correction to the conductance is a smooth function of the fraction of removed bonds, with $\sigma(q)= 1-2q+c_2q^2+\,\mathrm{higher}\,\,\mathrm{order}\,\,\mathrm{terms}\,\,\mathrm{in}\,\,\nicefrac{1}{L}$. This regular expansion indicates that, far from the percolation threshold, transport is governed by weak interactions between bond defects, and no anomalous exponents arise. The agreement between this perturbative form and numerical data over a wide range of $q$ underscores the robustness of this description in the weak-disorder regime.
	
	An interesting direction for future work is the study of spatial correlations in these networks. As seen in Fig.~\ref{fig:space}, the three models exhibit qualitatively different current distributions despite similar global conductance statistics. While our simulations are carried out only for $d=2$, we expect the qualitative features of our results to hold in higher dimensions as well.
	
	\section{Acknowledgments}
	\label{acknow}
	I am grateful to Deepak Dhar for his guidance, many insightful discussions, and a critical review of this manuscript. I am indebted to Samriddhi Sankar Ray for essential logistical support during my stay at ICTS-TIFR. I acknowledge the hospitality of ICTS-TIFR, Bengaluru and the support of the NIUS program of HBCSE-TIFR funded by the Department of Atomic Energy, Govt. of India (Project No. RTI4001). The simulations were performed on the ICTS clusters Mario and Tetris.

\end{document}